\documentclass[
	journal
]{IEEEtran}

\renewcommand{\markboth}[1]{\renewcommand{\leftmark}{#1}\renewcommand{\rightmark}{#1}}
\IEEEoverridecommandlockouts % This is to make the \thanks{} command visible in conference mode
\makeatletter
\def\markboth#1#2{\def\leftmark{\@IEEEcompsoconly{\sffamily}\MakeUppercase{\protect#1}}%
\def\rightmark{\@IEEEcompsoconly{\sffamily}\MakeUppercase{\protect#2}}}
\makeatother

\usepackage[amssymb]{SIunits}
\usepackage[cmex10]{amsmath}
\usepackage{amssymb}
\usepackage{amsthm}
\usepackage[latin1]{inputenc}
\usepackage{cite}
\usepackage{url}
\usepackage{epstopdf}
\usepackage{glossaries}
\usepackage{graphicx}
\usepackage[small,bf]{caption}
\usepackage[subrefformat=parens,labelformat=parens]{subfig}
\usepackage{bm}
\usepackage[usenames]{color}
\usepackage[lined,ruled,linesnumbered]{algorithm2e}
\usepackage{balance}
\usepackage{array}

\usepackage{mathtools}

\newcommand{\cA}{1}

\newenvironment{DIFnomarkup}{}{}

\newcolumntype{L}[1]{>{\raggedright\let\newline\\\arraybackslash\hspace{0pt}}m{#1}}
\newcolumntype{C}[1]{>{\centering\let\newline\\\arraybackslash\hspace{0pt}}m{#1}}
\newcolumntype{R}[1]{>{\raggedleft\let\newline\\\arraybackslash\hspace{0pt}}m{#1}}

\newcommand{\showproof}[1]{#1}
\newcommand{\showsketch}[1]{}

\newcommand{\EqC}{,}

\newcommand{\ph}{\tilde{p}}

\newcommand{\pow}[1]{^{(#1)}}
\newcommand{\expectation}[2]{\mathsf{E}_{#1}\left\{ #2 \right\}}
\newcommand{\Rho}{\mathrm{P}}

\mathtoolsset{showonlyrefs}

\graphicspath{%
	{../pstricks/}%
}%

\newacronym{AWGN}{AWGN}{additive white Gaussian noise} 
\newacronym{CSA}{CSA}{coded slotted ALOHA} 
\newacronym{ABCSA}{B-CSA}{all-to-all broadcast CSA} 
\newacronym{SNR}{SNR}{signal-to-noise ratio} 
\newacronym{SINR}{SINR}{signal-to-interference-plus-noise ratio}
\newacronym{PLR}{PLR}{packet loss rate} 
\newacronym{UEP}{UEP}{unequal error protection} 
\newacronym{BS}{BS}{base station} 
\newacronym{LDPC}{LDPC}{low-density parity-check} 
\newacronym{VN}{VN}{variable node} 
\newacronym{CN}{CN}{check node} 
\newacronym{DE}{DE}{density evolution} 
\newacronym{MDS}{MDS}{maximum distance separable} 
\newacronym{BEC}{BEC}{binary erasure channel} 
\newacronym{PEC}{PEC}{packet erasure channel} 
\newacronym{DAMA}{DAMA}{demand assignment multiple access} 
\newacronym{CSMA}{CSMA}{carrier sense multiple access} 
\newacronym{VANET}{VANET}{vehicular ad hoc network} 
\newacronym{V2V}{V2V}{vehicle to vehicle} 
\newacronym{PHY}{PHY}{physical} 
\newacronym{MAC}{MAC}{medium access control} 
\newacronym{ARQ}{ARQ}{automatic repeat request} 
\newacronym{CDMA}{CDMA}{code division multiple access} 
\newacronym{TDMA}{TDMA}{time division multiple access}
\newacronym{CAM}{CAM}{cooperative awareness message}
\newacronym{DENM}{DENM}{decentralized environmental notification message}
\newacronym{ETSI}{ETSI}{European Telecommunications Standards Institute}
\newacronym{GPS}{GPS}{Global Positioning System}
\newacronym{DUEP}{DUEP}{double unequal error protection}
\newacronym{VC}{VC}{vehicular communication}
\newacronym{RU}{RU}{receiving user}
\newacronym{SIC}{SIC}{successive interference cancellation}
\newacronym{EF}{EF}{error floor}
\newacronym{ID}{ID}{induced distribution}
\newacronym{RV}{RV}{random variable}
\newacronym{CDF}{CDF}{cumulative distribution function}
\newacronym{PMF}{PMF}{probability mass function}
\newacronym{PDF}{PDF}{probability density function}
\newacronym{iud}{i.u.d.}{independent and uniformly distributed}

\newcommand{\figref}[1]{Fig.~\ref{#1}}

\newcommand{\tabref}[1]{Table~\ref{#1}}

\newtheorem{proposition}{Proposition}

\newcommand{\setS}{\mathcal{S}}

\newcommand{\maxd}{d}

\begin{document}

\begin{DIFnomarkup}

\title{Unequal Error Protection in Coded Slotted ALOHA}

\author{
%Author 1, Author 2, Author 3, Author 4
Mikhail~Ivanov, Fredrik Br\"{a}nnstr\"{o}m, Alexandre Graell i Amat, and Gianluigi Liva
\thanks{This research was supported in part by the Swedish Research Council, under Grants No. 2011-5950 and 2011-5961, by the Ericsson's Research Foundation, by Chalmers Antenna Systems Excellence Center in the project `Antenna Systems for V2X Communication', and by the European Research Council, under Grant No. 258418 (COOPNET).
}
\thanks{M. Ivanov, F. Br\"{a}nnstr\"{o}m, and A. Graell i Amat are with the Department~of Signals and Systems, Chalmers University of Technology, SE-41296 Gothenburg, Sweden (e-mail: \{mikhail.ivanov, fredrik.brannstrom, \mbox{alexandre.graell}\}@chalmers.se).}
\thanks{G. Liva is with the Institute of Communications and Navigation (Digital
Networks Section) of the German Aerospace Centre (DLR), Oberpfaffen-
hofen, 82234 We{\ss}ling, Germany (e-mail: gianluigi.liva@dlr.de).}
}

\maketitle

\end{DIFnomarkup}

\begin{abstract} We analyze the performance of coded slotted ALOHA systems for a scenario where users have different error protection requirements {and correspondingly can be divided into user classes. The main goal is to design the system so that the requirements for each class are satisfied. To that end,} we {derive} analytical error floor approximations of the packet loss rate for each class in the finite frame length regime, as well as the density evolution in the asymptotic case. Based on this analysis, we propose a heuristic approach for the optimization of the degree distributions to provide the required unequal error protection. In addition, we analyze the decoding delay for users in different classes and show that better protected users experience a smaller average decoding delay. 
\end{abstract}

% Not needed for conference
% Redefine all acronyms that have been defined in the introduction
\glsresetall

\section{Introduction}\label{sec:intro}

\Gls{CSA} has recently been proposed as an uncoordinated \gls{MAC} protocol that can provide large throughputs~\cite{Paolini11,Stefanovic13}. Different types of \gls{CSA} have been proposed~\cite{Paolini14}. Most of them share a slotted structure borrowed from the original slotted ALOHA protocol~\cite{Roberts75} and the use of \gls{SIC}. The contending users introduce redundancy by encoding their messages into multiple packets, which are transmitted to the receiver in randomly chosen slots.

In this letter, we consider a framed \gls{CSA} system~\cite{Liva11}, where the messages of the users have different requirements in terms of the \gls{PLR}. We group the users into classes according to their requirements.\footnote{Dividing users into classes based on different channel conditions for frameless \gls{CSA} was used in~\cite{Stefanovic13d}.} 
For a standard (single-class) CSA, it was shown in~\cite{Ivanov16} that users with different repetition factors have different protection, i.e., \gls{CSA} inherently provides \gls{UEP}. We exploit this property by assigning different distributions to different classes. We refer to such a system as multi-class \gls{CSA}. {Assigning different distributions to different classes has been previously considered in~\cite{Toni15}, where multi-edge-type \gls{DE} was derived to analyze the asymptotic behavior of the system. In this letter, we first show that the DE in~\cite{Toni15} can be replaced by \emph{single-edge-type} DE based on the \emph{average} distribution.} We then extend the finite frame length \gls{EF} approximations in~\cite{Ivanov16} to the multi-class case. Based on the obtained analytical results, we propose a heuristic approach to the optimization of the degree distributions to satisfy specified \gls{PLR} requirements for each class in the finite frame length regime. Finally, we estimate the decoding delay for the obtained distributions.

\section{System Model}\label{sec:syst_model}

We consider the scenario where $m$ users transmit to a common receiver.\footnote{It is often assumed that the number of users is a Poisson random variable~\cite{Casini07}. In this letter, for simplicity we assume that $m$ is fixed.} {We assume that users are divided into $\kappa$ classes corresponding to different reliability requirements. A user belongs to  class $k$ with  probability~$\alpha_k$, such that $\sum_{k=1}^\kappa \alpha_k = 1$.} %{We furthermore assume that the average number of users in class 1 is smaller than that of class 2, i.e., $\alpha < 0.5$.} % The reliability requirements of the two classes will be discussed in~\secref{sec:optimization}.
Users transmit to the~receiver over a shared medium. The time is divided into \emph{frames}, consisting of $n$ slots of equal duration. We assume that all users are {frame- and slot-synchronized, which can be achieved by implementing a suitable signaling, as specified, e.g., in~\cite{dvb2011}. Each user transmits one message per frame by mapping it to a physical layer packet and repeating it $l$ times ($1 < l\le n$ is a random number selected according to a predefined distribution) in slots chosen uniformly at random within a frame.} %, as shown in~\figref{fig:system_model}. 
Every packet contains pointers to its copies, so that, once a packet is successfully decoded, full information about the location of the copies is available. A slot is called a singleton slot if it contains only one packet. If it contains more packets, we say that a collision occurs.

First, the receiver decodes the packets in singleton slots and obtains the location of their copies. {As soon as a packet is decoded, the channel coefficients associated with its copies can be accurately estimated~\cite{Casini07, Liva11}.} After subtracting the interference caused by the identified copies, decoding proceeds until no further singleton slots are found. We assume perfect channel estimation and interference cancellation, {as motivated by physical layer simulations in~\cite{Casini07, Liva11}.} 

The described system can be represented by a bipartite graph, in which users and slots are represented by \glspl{VN} and \glspl{CN}, respectively, and transmissions are represented by edges. A node is said to have the degree $l$ if it has $l$ incident edges. The decoding process is similar to the peeling decoding of low-density parity-check codes over the binary erasure channel~\cite{Liva11}.

We define the channel load as $g = m/n$.\footnote{{This definition coincides with the one of the instantaneous load in~\cite{Paolini14}. The results following from this definition can be applied to the case where $m$ is modeled as a random variable by averaging over the distribution of $m$.}} %The average number of users that successfully transmit their message, termed \emph{resolved users}, is denoted by $r$. The throughput $t = r/n$ shows how efficiently the frame is used. 
If the receiver fails to decode a packet from a user, we say that the user is unresolved. The average number of unresolved users in class $k$ is denoted by $\bar{r} \pow{k}$. In this letter, we analyze the \gls{PLR} for each user class as a function of $g$, defined as $p\pow{k}(g) = \bar{r} \pow{k} / \bar{m} \pow{k}$, where $\bar{m} \pow{k} = \alpha_k m$ is the average number of users in class~$k$. %{Since unresolved users do not retransmit their packets, the \gls{PLR} over  frames are independent. Hence, without loss of generality, we analyze the performance of the system on a frame base.}

The performance of the system greatly depends on the \gls{PMF} that users use to choose the degree $l$. This \gls{PMF} is referred to as the degree distribution and is specified in the form of a polynomial as
\begin{equation}\label{eq:distr_orig}
	\Lambda \pow{k}(x) = \sum_{l = 1}^{\maxd}\Lambda \pow{k}_{l}x^{l}\EqC
\end{equation}
where $\Lambda_l \pow{k}$ is the probability of choosing degree $l$ {for a user in class $k$} and $\maxd$ is the maximum degree. {Typically, $\maxd \ll n$.}

\section{Performance Analysis}

For later use, we define the average degree distribution as
\begin{equation}
\Lambda(x) =  \sum_{k = 1}^\kappa \alpha_k \Lambda \pow{k}(x).
\end{equation}

\subsection{Density Evolution}
For standard single-class \gls{CSA}, the \gls{PLR} performance exhibits a threshold behavior when $n \to \infty$, i.e., all but a vanishing fraction of the users are resolved if the channel load is below a certain threshold value{, denoted by $g^*$ and obtained by \gls{DE}~\cite{Liva11}.} %The threshold depends only on the degree distribution that users use to select their degree and it is 
Analogously, for multi-class CSA, we define $g_k^*$ to be the threshold for class-$k$ users, i.e., the largest channel load for which all but a vanishing fraction of class-$k$ users are resolved. The following proposition shows that the thresholds for all classes coincide and can be obtained by means of single-edge-type \gls{DE}.

\begin{proposition}\label{theor:uep_rx}
$g_1^* = g_2^*= \dots = g_\kappa^* = g^*$, where $g^*$ is the largest value of $g$ such that
\begin{equation}\label{eq:DE_eq}
\xi > 1 -  \exp{\left(-g\dot{\Lambda}(\xi)\right)}\EqC \quad \forall \xi \in (0, 1]\EqC
\end{equation}
where $\dot{\Lambda}(x) = \mathrm{d}\Lambda(x)/\mathrm{d}x$ denotes the derivative of $\Lambda(x)$.
\end{proposition}
\showproof{
\begin{IEEEproof}%[Proof outline]
We use the graph terminology and refer to users and slots as VNs and CNs, respectively. Let $\Rho\pow{k}(x) = \sum_{l} \Rho\pow{k}_l x^l$ denote the \gls{CN} degree distribution induced by class-$k$ VNs, i.e., $\Rho\pow{k}_l$ is the probability that $l$ class-$k$ VNs are connected to a given CN. Since the VNs select CNs at random, $\Rho\pow{k}_l$ is a Poisson distribution with mean $\dot{\Rho}\pow{k}(1)$~\cite{Liva11}. An edge connected to a class $k$ VN is called a type-$k$ edge. We define the edge-perspective \gls{VN} degree distribution for class $k$ as 
\[
\lambda\pow{k}(x) = \sum_{l = 1}^{\maxd} \lambda\pow{k}_l x^{l-1},
\]
where $\lambda\pow{k}_l$ is the probability that a type-$k$ edge is incident to a degree-$l$ VN. The edge-perspective \gls{CN} degree distribution for class $k$, denoted by $\rho\pow{k}(x)$, is similarly defined as
$\rho\pow{k}(x) = \sum_{l} \rho\pow{k}_l x^{l-1}$,
where $\rho\pow{k}_l$ is the probability that an edge is connected to a \gls{CN} with $l$ type-$k$ edges. It is easy to show that $\lambda\pow{k}(x) = \dot{\Lambda}\pow{k}(x)/ \dot{\Lambda}\pow{k}(1)$ and $\rho\pow{k}(x) = \dot{\Rho}\pow{k}(x)/ \dot{\Rho}\pow{k}(1)$. Moreover, using the definition of the Poisson distribution we can show that $\rho\pow{k}(x) = \Rho\pow{k}(x)$, i.e., the edge-perspective CN degree follows a Poisson distribution.

Let $q\pow{k}_i$ denote the probability that a class-$k$ VN is not resolved at the $i$th decoding iteration, i.e., $q\pow{k}_i$ is the probability that a class-$k$ VN sends an erasure to the \glspl{CN}. The probability that a selected CN sends an erasure back to a VN of class $k$, denoted by $\xi\pow{k}_i$, can then be expressed as
\begin{equation}\label{eq:proof}
	\xi\pow{k}_i \!= 1 - \expectation{l_1, \dots, l_\kappa}{\left(1 - q\pow{k}_i\right)^{l_k - 1}\!\prod_{\substack{j = 1\\ j\neq k}}^{\kappa} \left(1 - q\pow{j}_i\right)^{l_j}},
\end{equation}
where $l_k$ is the number of type-$k$ edges connected to the CN and $\expectation{}{\cdot}$ denotes expectation. Since $l_k$, $k = 1,\dots, \kappa$ are independent Poisson random variables, the probability in~\eqref{eq:proof} can be written as
\begin{equation*}
	\xi\pow{k}_i = 1 - \rho\pow{k}(1 - q\pow{k}_i) \prod_{\substack{j = 1\\ j\neq k}}^{\kappa} \Rho\pow{j}(1 - q\pow{j}_i).
\end{equation*}
Using the fact that $\rho\pow{k}(x) = \Rho\pow{k}(x)$, we obtain 
\begin{equation}\label{eq:proof2}
	\xi\pow{k}_i = 1 - \prod_{j = 1}^{\kappa} \Rho\pow{j}(1 - q\pow{j}_i).
\end{equation}
Since the right-hand side of~\eqref{eq:proof2} is independent of $k$, we conclude that $\xi\pow{1}_i = \xi\pow{2}_i = \dots = \xi\pow{\kappa}_i$ and to simplify the notation, we set $\xi\pow{1}_i = \dots = \xi\pow{\kappa}_i = \xi_i$. Since $\Rho\pow{k}_l$ is a Poisson distribution with mean $\dot{\Rho}\pow{k}(1)$, we can express $\xi_i$ as
\begin{equation}\label{eq:proof3}
\xi_i = 1 - \exp\left(- \sum_{k = 1}^{\kappa} \dot{\Rho}\pow{k}(1) q\pow{k}_i\right).
\end{equation}

A class-$k$ \gls{VN} sends an erasure back if all incoming messages are erased, which occurs with probability
\begin{equation}\label{eq:q}
q\pow{k}_{i+1} = \lambda\pow{k}(\xi_i).
\end{equation}
Substituting~\eqref{eq:q} into~\eqref{eq:proof3} gives
\begin{align}
\xi_i &= 1 - \exp\left(- \sum_{k = 1}^{\kappa} \dot{\Rho}\pow{k}(1) \lambda\pow{k}(\xi_{i-1})\right) \\
&= 1 - \exp\left(- \sum_{k = 1}^{\kappa} \frac{\dot{\Rho}\pow{k}(1)}{\dot{\Lambda}\pow{k}(1)} \dot{\Lambda}\pow{k}(\xi_{i-1})\right).\label{eq:proof_xi_i}
\end{align}
Since the number of type-$k$ edges is $n \dot{\Rho}\pow{k}(1) = \alpha_k m \dot{\Lambda}\pow{k}(1)$, we have
\begin{equation}
\frac{\dot{\Rho}\pow{k}(1)}{\dot{\Lambda}\pow{k}(1)} = \frac{\alpha_k m}{n} = g \alpha_k .
\end{equation}
Inserting the obtained result into~\eqref{eq:proof_xi_i} gives
\begin{align}
\xi_i &= 1 - \exp\left(- g \sum_{k} \alpha_k \dot{\Lambda}\pow{k}(\xi_{i - 1})\right) \\
&=  1 - \exp\left(- g \dot{\Lambda}(\xi_{i - 1})\right),
\end{align}
where the last step follows from the linearity of the derivative. Hence, $\xi_i$ tends to zero as $i \to \infty$ if the condition in~\eqref{eq:DE_eq} is satisfied.

Finally, the probability of a class-$k$ VN to be erased at the $i$th decoding iteration is given by
\begin{equation}
	p\pow{k}_i = \Lambda\pow{k}(\xi_i).
\end{equation}
Hence, $p\pow{k}_i > 0$ for $k = 1,\dots,\kappa$ if $\xi_i>0$ and  $p\pow{k}_i = 0$ for $k = 1,\dots,\kappa$ if $\xi_i = 0$, which proves that the threshold is the same for all classes.
\end{IEEEproof}
}
\showsketch{
\begin{IEEEproof}[Proof outline]
The proposition can be proved by writing two multi-edge-type \gls{DE} equations for each class relating the erasure probabilities for the messages exchanged between \glspl{VN} and \glspl{CN}. Since each class of users induces a Poisson degree distribution on the \glspl{CN}, it is possible to show that the erasure probability at the output of the \glspl{CN} is independent of the class of the \gls{VN} to which the message is sent. Thus, single-edge-type \gls{DE} is sufficient for determining the threshold.\footnote{A detailed proof can be found in an extended version of this letter~\cite{Ivanov16j}.}
\end{IEEEproof}
}
\begin{samepage}
We remark that Proposition 1 is also valid if the fraction of users in each class is fixed as in~\cite{Toni15}. It is easy to verify that~\cite[eq.~(18)]{Toni15} can be reduced to~\eqref{eq:DE_eq}.
Since the thresholds for different classes coincide, we conclude that \gls{DE} provides little support for the design of multi-class CSA systems and one needs to look at the finite frame length performance instead.
\subsection{Packet Loss Rate}
The finite frame length regime gives rise to an \gls{EF} in the PLR performance of 
\end{samepage}
\gls{CSA}, i.e., the \gls{PLR} is bounded away from zero even for channel loads below the threshold. The \gls{EF} is caused by harmful structures in the corresponding bipartite graph, commonly referred to as stopping sets~\cite{Ivanov16}. A connected bipartite  graph $\setS$ is said to be a stopping set if all check nodes in $\setS$ have a degree larger than one. The analysis of stopping sets allows to accurately predict the \gls{EF} for standard \gls{CSA}~\cite{Ivanov16}. In this letter, we extend these results to multi-class \gls{CSA}.

Let $p_l (g)$ denote the probability that a degree-$l$ user is not resolved by the receiver. {We can then easily derive the \gls{PLR} for class-$k$ users as
\begin{equation}\label{eq:plr_class}
p \pow{k} (g) =  \sum_{l = 0}^{d}\Lambda \pow{k}_{l} p_{l} (g),
\end{equation}
which can be seen as a generalization of~\cite[eq.~(10)]{Ivanov16} to the multi-class case.} An approximation of $p_l (g)$ in the \gls{EF} region can be obtained as~\cite{Ivanov16}
\begin{multline}\label{eq:final_aprx}
	p_l (g) \approx \frac{(m-1)!}{\Lambda_l} \\ \times \sum_{\setS \in \mathcal{A}}{ \frac{v_{l}(\setS) c(\setS)}{(m - \nu(\setS))!} \binom{n}{\mu(\setS)}\prod_{j = 1}^{d}{\binom{n}{j}^{-v_j(\setS)}\frac{\Lambda_{j}^{v_{j}(\setS)}}{v_j(\setS)!}}}\EqC
\end{multline}
where $\mathcal{A}$ is the set of the considered stopping sets, $\nu(\setS)$ and $\mu(\setS)$ are the number of variable and check nodes in $\setS$, respectively, $v_j(\setS)$ is the number of degree-$j$ variable nodes in $\setS$, and $c(\setS)$ is the number of graphs isomorphic with $\setS$. 

We remark that the analysis in~\cite{Ivanov16} is based on enumerating stopping sets and is able to predict the \gls{EF} for $l \le 4$ at low-to-moderate channel loads. To predict the \gls{EF} for larger $l$, the set of considered stopping sets needs to be extended.

\section{Distribution Optimization}\label{sec:optimization}
In this section, we provide an efficient method to design good distributions for {multi-class \gls{CSA} which satisfy specified \gls{PLR} constraints (for a given $n$, $g$, and $\alpha_k$, $k = 1, \dots, \kappa$). For simplicity, we consider  only two classes of users.} {Ideally, we would like to design distributions that provide the required level of reliability at the highest possible channel load. Unfortunately, the analytical \gls{EF} approximation is inaccurate for large channel loads. Hence, we resort to a heuristic optimization based on the threshold and the \gls{EF} approximation, which was shown to be useful in~\cite{Ivanov16}.} {Without loss of generality, we assume that class $\cA$ requires higher reliability.}

The \gls{PLR} requirements for class $k = 1,2$ are described by the target PLR $\ph\pow{k}$ at a specific channel load $\tilde{g}$. To reduce the search space, we restrict the distributions to be in the form $\Lambda \pow{k}(x) = \Lambda_2 \pow{k} x^2 + \Lambda_3 \pow{k} x^3 + \Lambda_8 \pow{k} x^8$. By limiting to these degrees, it was shown in~\cite{Liva11} that large thresholds can be obtained, while reasonably low \gls{EF} can be provided by carefully controlling the fraction of degree-$2$ users~\cite{Liva11}. We numerically solve the following optimization problem
\begin{equation}\label{eq:joint_opt}
\begin{aligned}
& \underset{\Lambda\pow{1}(x), \Lambda\pow{2}(x)}{\text{maximize}}
& &  g^*\\
& \text{\hspace{0.25cm}subject to}& & p \pow{k}(\tilde{g}) \le \ph\pow{k} \text{ for } k = 1,2\\
\end{aligned}
\end{equation}
by means of the Nelder-Mead simplex algorithm~\cite{Lagarias98}. We use~\eqref{eq:plr_class} together with the approximation~\eqref{eq:final_aprx} to estimate the PLR for the two classes. We remark that~\eqref{eq:final_aprx} gives an accurate prediction for low-to-medium channel loads, hence, $\tilde{g}$ should not be too large to make the optimization problem meaningful. The employed optimization algorithm is highly sensitive to the initial value of the distributions. We solve this problem by uniformly sampling the search space with a step $0.1$ and running the optimization multiple times. 

The results of the optimization for two values of $\alpha_1$ and different values of $\ph \pow{k}$ are presented in~\tabref{tab:opt_distr}. The frame length is set to $n = 100$ and the target channel load is $\tilde{g} = 0.5$. It can be seen that the larger $\alpha_1$, the smaller the threshold since the average PLR requirements are stricter. We can also observe that in most of the cases $\Lambda \pow{1}_2 \approx 0$ and $\Lambda \pow{1}_8 \approx 1$ to guarantee a low error floor, whereas $\Lambda \pow{2}_8$ is close to zero to keep the threshold reasonably large. The main goal of the optimization is therefore to carefully select parameters $\Lambda \pow{2}_2$ and $\Lambda \pow{2}_3$.

\begin{table}
\caption{Optimized distributions for $n =100$ and different values of $\alpha_1$ and target PLR at $\tilde{g} = 0.5$.}
\begin{center}

\subfloat[$\alpha_1 = 0.1$.]{
  \begin{tabular}{c|c|c|c|c|c|c|c|c}
	  \hline
	$\ph \pow{1}$ & $\ph \pow{2}$ & $\Lambda \pow{1}_2$ &$\Lambda \pow{1}_3$&$\Lambda \pow{1}_8$&$\Lambda \pow{2}_2$&$\Lambda \pow{2}_3$&$\Lambda \pow{2}_8$& $g^*$\\
	\hline
	$10^{-5}$ &$10^{-2}$ & 0& 0.01 & 0.99& 0.57 & 0.30& 0.13 & 0.94\\
	\hline	   
	$10^{-4}$ &$10^{-3}$& 0.02 & 0.11& 0.87 & 0.25& 0.66 & 0.09 & 0.89\\
	\hline	    
	$10^{-5}$ &$10^{-3}$& 0& 0.01 & 0.99 & 0.25 & 0.67& 0.08  & 0.89\\
	\hline	    
		$10^{-5}$ &$10^{-4}$ &0.01 &0 &0.99 & 0.04 & 0.51 & 0.45 & 0.72\\
	\hline	    
  \end{tabular}

}

\subfloat[$\alpha_1 = 0.2$.]{
  \begin{tabular}{c|c|c|c|c|c|c|c|c}
	  \hline
	$\ph \pow{1}$ & $\ph \pow{2}$ & $\Lambda \pow{1}_2$ &$\Lambda \pow{1}_3$&$\Lambda \pow{1}_8$&$\Lambda \pow{2}_2$&$\Lambda \pow{2}_3$&$\Lambda \pow{2}_8$& $g^*$\\
	\hline
%	$10^{-2}$ &$10^{-3}$& 0.64 & 0.37& 0.02 & 0& 0 & 0.99 & 0.94	\\	
%		\hline
%	$10^{-2}$ &$10^{-4}$&  & & & &  & & \\	
%	\hline
	$10^{-5}$ &$10^{-2}$& 0& 0.01 & 0.99 & 0.64 & 0.33& 0.03  & 0.94\\
	\hline	    
	$10^{-4}$ &$10^{-3}$ & 0& 0.25 & 0.75 & 0.26 & 0.72& 0.02 & 0.89\\	
	\hline	    
	$10^{-5}$ &$10^{-3}$ & 0& 0.01 & 0.99 & 0.27 & 0.73& 0 & 0.88\\
		\hline	    
		$10^{-5}$ &$10^{-4}$ & 0.02& 0.02& 0.96 & 0 & 0.63& 0.37 & 0.72\\
	\hline	    

  \end{tabular}
}  

%\subfloat[$\alpha_1 = 0.4$.]{
%  \begin{tabular}{c|c|c|c|c|c|c|c|c}
%	  \hline
%	$\ph \pow{1}$ & $\ph \pow{2}$ & $\Lambda \pow{1}_2$ &$\Lambda \pow{1}_3$&$\Lambda \pow{1}_8$&$\Lambda \pow{2}_2$&$\Lambda \pow{2}_3$&$\Lambda \pow{2}_8$& $g^*$\\
%	\hline
%		$10^{-5}$ &$10^{-2}$ & 0& 0.03 & 0.97 & 0.95 & 0.05& 0 & 0.87\\
%	\hline	    
%	$10^{-4}$ &$10^{-3}$ & 0& 0.34 & 0.66 & 0.32 & 0.68& 0 & 0.86\\
%	\hline	    
%	$10^{-5}$ &$10^{-3}$& 0& 0.04 & 0.96 & 0.32 & 0.68& 0  & 0.80\\
%	\hline	    
%		$10^{-5}$ &$10^{-4}$ & 0 & 0.03 & 0.97 & 0.04 & 0.65 & 0.31 & 0.69 \\
%	\hline	    
%
%  \end{tabular}
%}
  \end{center}
  \label{tab:opt_distr}
  \vspace{-0.7cm}
\end{table}

\begin{figure}
\centering
\vspace{-0.4cm}
	\subfloat[]{
		\includegraphics[]{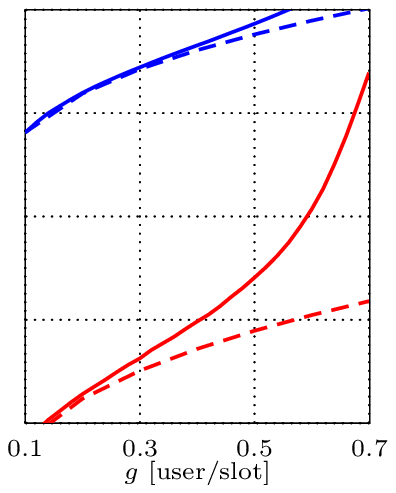}\label{fig:ex3}
	}
	\subfloat[]{
		\includegraphics[]{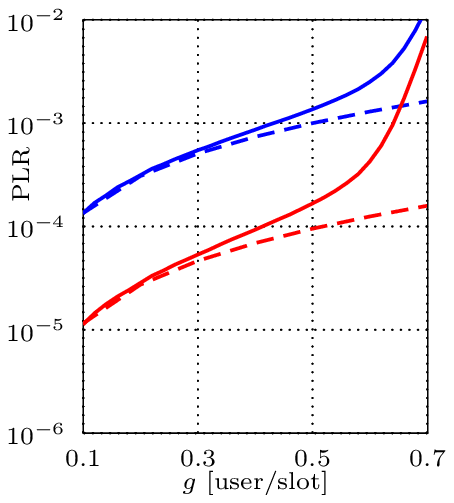}\label{fig:ex1}
	}
	\vspace{-0.2cm}
	\subfloat[]{
		\includegraphics[]{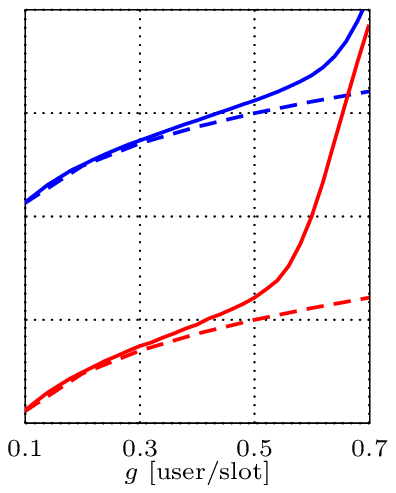}\label{fig:ex2}
	}
	\subfloat[]{
		\includegraphics[]{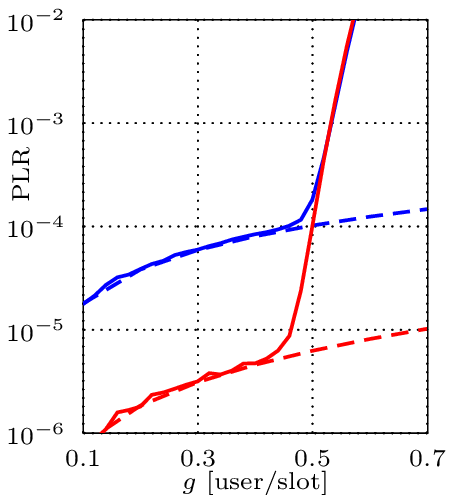}\label{fig:ex4}
	}
	\caption{PLR performance for the optimized distributions in~\tabref{tab:opt_distr}(b) for $\alpha_1 = 0.2$ and $n = 100$. Red and blue curves show the performance for the class 1 and class 2, respectively. Solid lines show simulation results. Dashed lines show the \gls{EF} approximation.}
	 \vspace{-0.4cm}
	\label{fig:sim_r14}
\end{figure}

The performance of the optimized distributions for $\alpha_1 = 0.2$ is shown in~\figref{fig:sim_r14}. Solid lines show simulation results and dashed lines show the analytical \gls{EF} predictions in~\eqref{eq:plr_class} based on~\eqref{eq:final_aprx}. We note that the PLR requirements are met by the analytical approximations. However, the actual performance violates the requirements in most of the cases, hence, $\tilde{g}$ needs to be selected with this consideration in mind. 

It can also be seen from the figures that for low target PLRs (i.e., relatively low values of the threshold) the simulated curves have a more pronounced waterfall and agree well with the analytical curves in the \gls{EF} region. From~\figref{fig:sim_r14}(a) we can also see that relaxed PLR requirements result in a higher threshold. However, the mismatch between simulations and analytical results increases in this case. Hence, the actual performance is farther away from the target PLR. Similar effect is observed when the requirements are very strict and the waterfall begins at channel loads lower than $\tilde{g}$, as in \figref{fig:sim_r14}(d).  {We also remark that the waterfall regions of the PLR curves for different classes start at approximately the same channel load, which confirms Proposition~1. The difference between the waterfall regions (especially pronounced in~\figref{fig:sim_r14}(a) due to a high threshold) is caused by the finite frame length regime.} Simulation results for other values of $\alpha_1$ and $n$ (not included here) show very similar behavior.

\section{Decoding Delay}

In this section, we analyze the decoding delay for users from different classes. When considering framed versions of CSA it is usually assumed that the receiver first receives the entire frame and then performs decoding. Assuming further that signal processing does not introduce any delay, i.e., decoding, channel estimation, and \gls{SIC} are instantaneous, the decoding delay is equal to the frame duration if a user is resolved and it is not defined otherwise.

Here, to provide faster decoding, we use the slot-based decoder proposed in~\cite{Stefanovic13} for frameless CSA. In frameless CSA~\cite{Stefanovic13}, it is essential that the receiver attempts decoding continuously on a slot-by-slot basis since the contention needs to be terminated once a certain criterion is satisfied. We can use this decoder also in our scenario to provide faster decoding of the users. The decoding delay, denoted by $\Delta t$, is then a \gls{RV} that is equal to the number of slots the receiver needs to observe to decode a user. For convenience, we normalize it by the frame length.

Predicting analytically the decoding delay of the slot-based decoder is difficult in general. Obviously, it can be upperbouded by the frame duration. Moreover, the delay can be easily analyzed at asymptotically low channel loads, i.e., when $g \to 0$. To that end, we assume that a single user is present in the system and it selects degree $l$ according to a degree distribution. The decoding delay for this user is denoted by $d_l$ and it is equal to the smallest slot number that the user chooses for transmission normalized by the frame length $n$. Assuming that $n \to \infty$, the distribution of $d_l$, denoted by $f(d_l)$, is the distribution of the minimum of $l$ independent and uniformly distributed on $[0,\,1]$ \glspl{RV} and it is given by
\begin{equation}
f (d_l)= 
\begin{cases}
l(1 - d_l)^{l-1}& \text{ for } 0\le d_l\le 1,\\
0 & \text{ otherwise}.
\end{cases}
\end{equation}
The mean of $d_l$ for a degree-$l$ user is $\bar{d}_l = (l+1)^{-1}$. The average delay for class $k$ at asymptotically low channel load, denoted by $\bar{d} \pow{k}$, can be calculated as
\begin{equation}
\bar{d} \pow{k} = \sum_{l = 1}^{d} \Lambda_l \pow{k} \bar{d}_l.
\end{equation}

Finding the distribution of the delay for $g>0$ is not easy and should take collisions into account.  Their effect is twofold. In the \gls{EF} region, nearly all collisions are resolvable, and hence, they only postpone decoding of the colliding users. At high channel loads, colliding users create stopping sets and these users are excluded from the delay estimation, which results in changing the statistics of the delay. {We therefore resort to simulations to estimate the decoding delay for channel loads larger than zero.}

\begin{figure}
	\includegraphics{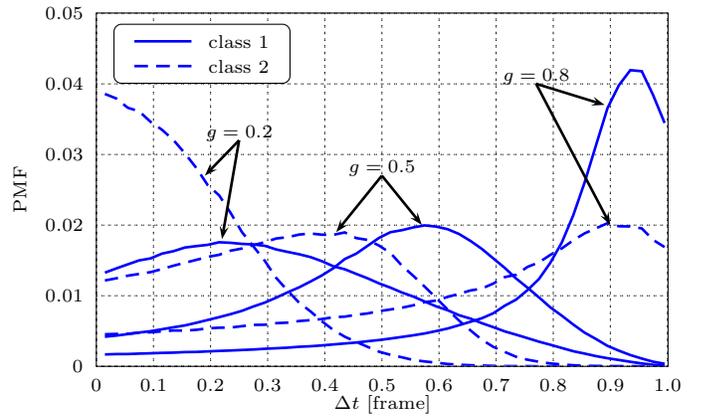}
	\caption{{PMF of the delay for the second pair of distributions in~\tabref{tab:opt_distr}(b) and $n = 100$ for different channel loads. Solid and dashed lines correspond to class 1 and class 2, respectively.}}
	\label{fig:delay_pdf}
\end{figure}

\begin{figure}
	\includegraphics{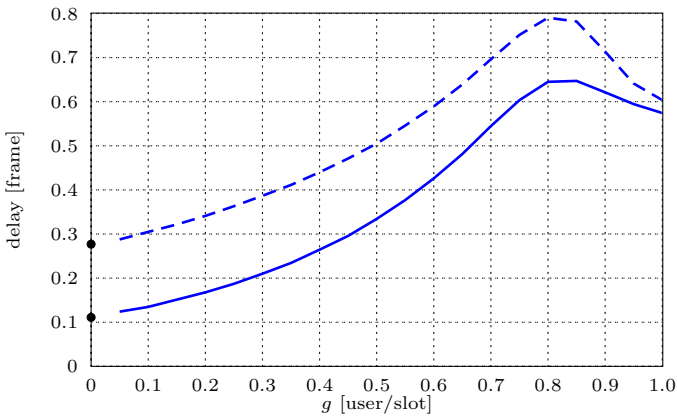}
	\caption{{Average delay as a function of the channel load for the second pair of distributions in~\tabref{tab:opt_distr}(b) and $n = 100$. Solid and dashed lines correspond to class 1 and class 2, respectively.}}
	\label{fig:delay_mean}
	\vspace{-0.3cm}
\end{figure}

The estimated \glspl{PMF} of the delay for the second pair of distributions in~\tabref{tab:opt_distr}(b) ($n = 100$) and different values of the channel load are shown in~\figref{fig:delay_pdf}. In~\figref{fig:delay_mean}, we show the estimated average delay as a function of the channel load. It is observed that class-1 users experience faster decoding than class-2 users. For example, at $g = 0.5$ class-1 users are resolved after $0.33n$ slots on average, whereas class-2 users are resolved after observing half a frame. However, the difference disappears as $g \to 1$. It is worth noting that only resolved users are accounted for when calculating the delay. The decoding delay for $g > g^{*}$ is not of interest since very few users are resolved. The analytical delays $\bar{d} \pow{1} = 0.27$ and $\bar{d} \pow{2} = 0.11$ when $g \to 0$ are shown with black dots and they agree well with the simulation results. 
The simulation results also indicate that the gap between the delay curves for different classes is consistent in the low-to-moderate load region, rendering the analytical results useful even for $g>0$.

\section{Conclusions}
In this letter, we analyzed the performance of a multi-class CSA system where different classes of users, which have different error rate requirements, are assigned different distributions. We presented a framework to design degree distributions for finite frame lengths to provide \gls{UEP} based on \gls{EF} approximations and asymptotic thresholds. We further analyzed the decoding delay for different classes of users when using a slot-based decoder. The results show that multi-class CSA is capable of providing different levels of protection  at high channel loads, as well as a smaller average decoding delay for better protected users.

\balance


\end{document}